\documentclass{aa}
\usepackage{graphicx,times}
\newlength{\fsize}
\newcommand{\mk}{}

\def\Real{{\rm I\mathchoice{\kern-0.70mm}{\kern-0.70mm}{\kern-0.65mm}%
  {\kern-0.50mm}R}}  
  \def\bx#1{\leavevmode\thinspace\hbox{\vrule\vtop{\vbox{\hrule\kern1pt
  \hbox{\vphantom{\tt/}\thinspace{\bf#1}\thinspace}}
  \kern1pt\hrule}\vrule}\thinspace}

\def\be{\begin{equation}} \def\ee{\end{equation}}

\begin{document}

\title{Evolution of the magnetic field in magnetars}

\author{J.~Braithwaite \inst{1,2}\and H.~C.~Spruit \inst{1,3}}
\institute{$~^1$  Max-Planck-Institut f\"ur Astrophysik, Karl-Schwarzschild-Stra{\ss}e 1,
Postfach 1317,  D--85741 Garching, Germany\\
\tt$^2$jon@mpa-garching.mpg.de, $^3$henk@mpa-garching.mpg.de}
\authorrunning{Braithwaite and Spruit}
\titlerunning{Evolution of the magnetic field in magnetars}

\abstract{We use numerical MHD to look at the stability of a possible
poloidal field in neutron stars (\cite{FloandRud:1977}), and follow
its unstable evolution, which leads to the complete decay of the field. We
then model a neutron star after the formation of a solid crust of high
conductivity. As the initial magnetic field we use the stable `twisted
torus' field which was the result of our earlier work
(\cite{BraandNor:2005}),
since this field is likely to exist in the
interior of the star at the time of crust formation. We follow the
evolution of the field under the influence of diffusion, and find that
large stresses build up in the crust, which will presumably lead to
cracking. We put this forward as a model for outbursts in soft gamma
repeaters.
\keywords {instabilities -- magnetohydrodynamics (MHD) -- stars:
magnetic fields -- magnetars}}

\maketitle

\section{Introduction}
\label{sec:intro}

The {\it soft gamma repeaters} (SGRs) are characterised by continuous
emission of X-rays of luminosity $10^{35} - 10^{36}\,\rm{erg\,s}^{-1}$
(\cite{Rothetal:1994}, \cite{Hurleyetal:1999}) and X-ray outbursts, each
lasting less than a second, but in total accounting for a similar
luminosity to the continuous emission. These outbursts are extremely
bright ({\mk up to} $10^{42}\,\rm{erg\,s}^{-1}$) and therefore super-Eddington ($10^4
L_{\rm{Edd}}$), and in {\mk three} of these
objects (the whole class contains only four known specimens) a much
brighter outburst has been observed. In SGR 0526-66 an outburst on 5th
March 1979 released X-rays to the tune of an estimated
$4\times10^{44}\,\rm{erg}$, SGR 1900+14 released
$1\times10^{44}\,\rm{erg}$ in its 27th August 1998 outburst (\cite{Cline:1982},
\cite{Mazetsetal:1999}), {\mk while the most powerful event of all, on 27th December 2004, SGR 1806-20 produced an estimated $2\times10^{46}\,\rm{erg}$ (\cite{Palmeretal:2005})}. These giant
outbursts consisted of a short intense burst of a fraction of a
second, followed by an afterglow lasting {\mk several} minutes.
Regular variability in the X-ray flux from these objects has been
detected, which can most easily be interpreted as the rotational
period, $P$. The periods range from $5$ to $8$ seconds. If one measures the
period over a long period of time, one can measure the rate of change
of period $\dot{P}$ and hence calculate a {\it characteristic age}
$P/\dot{P}$. For two SGRs, values of around $3000$ years have been
measured (\cite{Kouveliotouetal:1998}, \cite{Woodsetal:2000}).

{\mk About 8} further objects have been classified as {\it anomalous X-ray pulsars}
(AXPs) -- their observational characteristics are essentially those of
SGRs without outbursts. They emit X-rays at the same luminosity,
rotate at the same speed ($6$ to $12$ seconds) and have the same
characteristic ages. This similarity suggests that AXPs are nothing
but SGRs in some kind of dormant state (\cite{ThoandDun:1996},
\cite{Mereghetti:2000}). 

That most of these objects have been associated with young supernova
remnants leads naturally to the hypothesis that they are neutron
stars. The question of the nature of the energy source for their X-ray
emission has been answered in several ways, although most authors now
agree on a strong decaying magnetic field, the {\it magnetar}
model, first proposed by Duncan \& Thompson (1992). According to this
model, the observed X-ray luminosity (in quiescence and in outburst)
is powered by the release of magnetic energy in a very strong
($B>10^{14} \rm{G}$) field. In contrast, the accretion and fossil-ring models have
run into significant difficulties in recent years.

The evidence for such a strong field is the rapid rotational spindown
of these objects (see, for example, {\mk Shapiro \& Teukolsky (1983)}
for an explanation of the relevant physics). A magnetic field of the inferred
strength will contain about $10^{47}$ erg -- enough to account for the
observed luminosity for around $10^4$ years. This is significantly
more energy than these stars contain in the form of rotational kinetic
energy -- a neutron star rotating with a period of $6$ seconds will
contain just $5\times10^{44}$ erg. The spindown luminosity (the rate
of change of this rotational kinetic energy) of SGRs is measured at
$10^{34}\,\rm{erg\,s}^{-1}$, between one and two orders of magnitude too
small to account for the X-ray luminosity. These objects can
therefore, unlike classical pulsars, not be powered by spindown alone.
This is the main motivation for the magnetic decay model for SGR/AXP
emission.

In this paper, we develop the magnetic decay model a little further by presenting 
stable field configurations obtained from 3D numerical simulations, and
by computing the pattern of crustal stress that develops when such a field
evolves by diffusion. 

\subsection{Powering by magnetic field decay}
In the magnetic decay model
the primary cause of the decay is energy release by some form of rearrangement of the 
magnetic field configuration in the star. Most of the magnetic energy is
contained in the interior, and a smaller (but possibly comparable) amount 
in the atmosphere. Rearrangements therefore release energy both in the 
atmosphere and in the interior. To the extent that this is a slow continuous process
(`creep'), it powers quiescent emission, both in thermal form (heating of the
interior) and non-thermal (the energy released in the atmosphere). Apparently,
some rearrangement takes place suddenly (`fracture'). The atmospheric energy 
release in such an event powers the observed outbursts while the internally 
dissipated energy leaks out more slowly, adding to the quiescent emission 
(\cite{ThoandDun:1996}). A straightforward extension of this interpretation is 
the possibility that the outburst episodes actually involve cracking of the 
neutron star crust and consequent release of magnetic energy in the atmosphere 
(\cite{ThoandDun:1995}). A slow steady change in the interior field, with the surface field kept 
in place by the solid crust, would slowly build up magnetic stress at the 
crust/core boundary, which  is released in crustal quakes. 

\subsection{Origin of the field}
The magnetic fields of neutron stars were probably already present at birth
(the possibility of a later origin by thermomagnetic effects seems less likely,
cf.\ Blandford et al. 1983). One possibility could be convection in the 
proto-neutron star, when unstable density gradients are produced by neutrino
cooling. Possible field strengths up to $10^{15}$ G may be generated by a 
convective dynamo (e.g. \cite{ThoandDun:1993}). Another possibility is that the magnetic
field is formed by compression during core collapse, from a field
already existing in the progenitor star. 

In either case, the magnetic field formed is likely to be out of equilibrium and
unstable directly after the collapse, raising the question how much of the
magnetic field can survive in the subsequent evolution when the star is still 
fluid. Once a solid crust has formed we may assume that at least the surface field 
is frozen, with further evolution taking place only on a much longer
diffusive time-scale. It is thought that the crust will not begin to form until
around $100$ s after 
the collapse. If the field in the interior  is of order $10^{15}$ gauss, the Alfv\'{e}n 
crossing time will be of order $t_{\rm A}=\sqrt{4\pi \rho}
R_\ast/B\approx 0.1$ s. 
Thus any unstable field has had ample time to either evolve, either into a stable 
configuration of lower energy (if this exists) or decay to nothing, before it 
can be `frozen in' by the crust.

This estimate is modified a little when we take into account that the star is
likely to be rapidly rotating, with a period $P$ which may be shorter than 
the Alfv\'en crossing time $t_{\rm A}$. In this case, the magnetic instability 
time-scale is lengthened by a factor $t_{\rm A}/P$. For a plausible rotation
period of $10$ ms, the instability time-scale is then increased to
$\approx 1$ s, still short compared with the time till crust formation.

An important question is thus what happens to an initially unstable field 
configuration, and if/how it can evolve into a stable form. 












\section{Stable and unstable field configurations}
\label{sec:stableandunstable}

\begin{figure}
\includegraphics[width=1.0\hsize,angle=0]{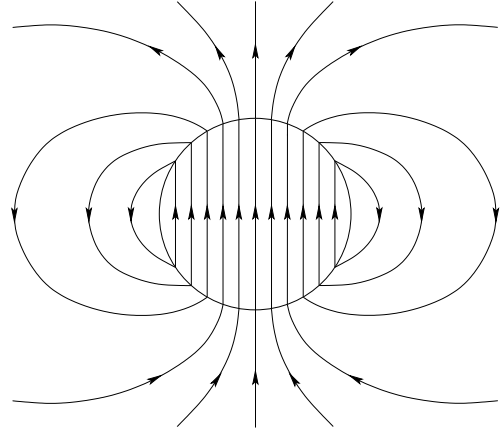}
\caption{The Flowers-Ruderman field.}
\label{fig:initfield}
\end{figure}

\begin{figure*}
\includegraphics[width=1.0\hsize,angle=0]{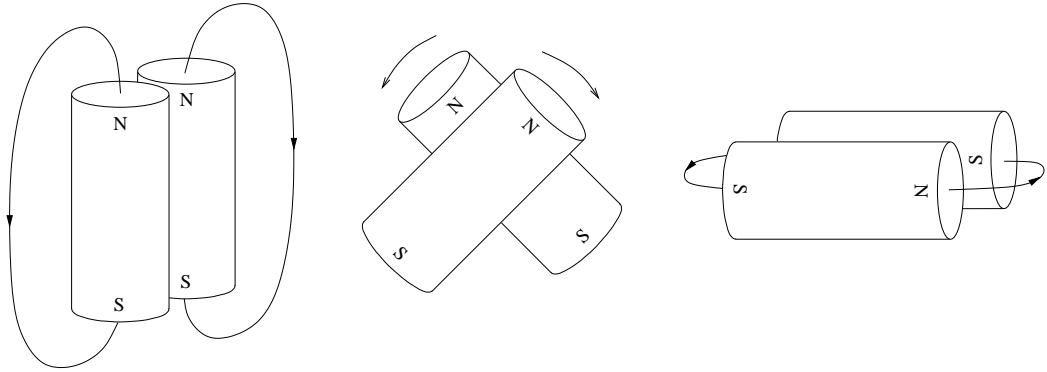}
\caption{{\mk Flowers-Ruderman argument for the instability of a star with a uniform interior magnetic field.} Two bar magnets are free to rotate about a common axis. If
pointing in the same direction, they are unstable and will rotate
until pointing in opposite directions.}
\label{fig:barmagnets}
\end{figure*}

We are interested in finding a field configuration which could account
for the behaviour of these neutron stars. It is unclear what
kind of field a neutron star may be born with, or rather, what kind of
field it will have when the crust has formed. 
In a differentially rotating star, purely
toroidal field configurations (azimuthal fields) form naturally from a weak initial
field by winding up of field lines (\cite{Spruit:2002} and references therein). The proto-neutron star may well be a 
strongly differentially rotating object that would nicely wind up field lines. It is 
known, however, that predominantly toroidal fields are all unstable on an Alfv\'en 
crossing time (\cite{Tayler:1973}, \cite{Acheson:1978}, \cite{PitandTay:1986},
\cite{Braithwaite:2005}). 

At the other extreme, one can think of a purely poloidal field (field lines
confined to meridional planes, $B_\phi=0$).  Such a field is again known to
be unstable on an Alfv\'en crossing time (\cite{Wright:1973}, Markey
\& Tayler 1973, 1974) if it has field lines that are 
closed inside the star. If the field lines close only outside the star, the stability 
analysis is slightly different but the result the same. A
well-known simple example of this case is a configuration consisting of a uniform
field inside the star, with a matching dipole outside the star (Fig. \ref{fig:initfield}).
An argument by Flowers and Ruderman shows why this field is unstable in the
absence of a crust. Since this example has played some role in the discussion of
neutron star magnetic fields, we study it in some detail below, with a numerical
simulation in Sect. \ref{sec:instdip}. 

\subsection{An unstable field configuration}
\label{sec:unstable}

Flowers and Ruderman (1977) consider a uniform field in the neutron
star interior with a potential field outside it. They find this to be
unstable on a dynamic time-scale (Alfv\'{e}n crossing time-scale).
A simple way to understand this (see Fig. \ref{fig:barmagnets})
is to imagine the uniform field as a collection of parallel bar magnets.
Two bar magnets side by side, pointing in the same
direction will repel each other, since the north pole of one
will be next to the north pole of the other. They will tend to rotate
so that the north pole of one is next to the south pole of the other,
reducing the magnetic energy outside but not inside the magnets.
In a fluid star, without stabilising solid-state forces, the equivalent
process would be to cut the star in half along a plane along the field
lines, and rotate one of the halves by  $180^{\circ}$. This reduces
the energy in the potential field outside the star in the same way as 
in the case of the bar magnets, whilst all energy forms inside the star 
(magnetic, thermal and gravitational) remain unchanged, also as in 
the case of the bar magnets. 

Flowers and Ruderman then consider ways in which a field may be stabilised.
Firstly, the solid crust of the neutron star, if it forms quickly enough,
could prevent the decay of a field. Secondly, the addition of a
toroidal component could stabilise the field, because its energy would
necessarily {\it increase} during the motion depicted in Fig.
\ref{fig:barmagnets}. To understand this, imagine wrapping an elastic
band around the two bar magnets. Rotating the magnets now will stretch
the elastic band, requiring energy. Field lines behave in a similar
way, because to stretch them also requires energy. 

\subsection{Stable field configurations and the role of helicity}
\label{sec:evoltorus}

It was predicted by Prendergast (1956) that a stable field inside a
star may consist of a polar dipole component stabilised by a toroidal
component of comparable strength. This principle is applied with success in
the design of fusion reactors. Recently it has become possible to
follow the evolution of an arbitrary field with numerical
magnetohydrodynamics.
This was done by Braithwaite \& Nordlund (2005, hereafter paper I -- {\mk for a summary of the results see \cite{BraandSpr:2004} -- where the related problem of the origin and evolution of the fields
of magnetic A stars was studied.}
The main result from this work is the emergence of a stable 
field configuration, independent of the initial field configuration. 

\begin{figure}
\includegraphics[width=1.0\hsize,angle=0]{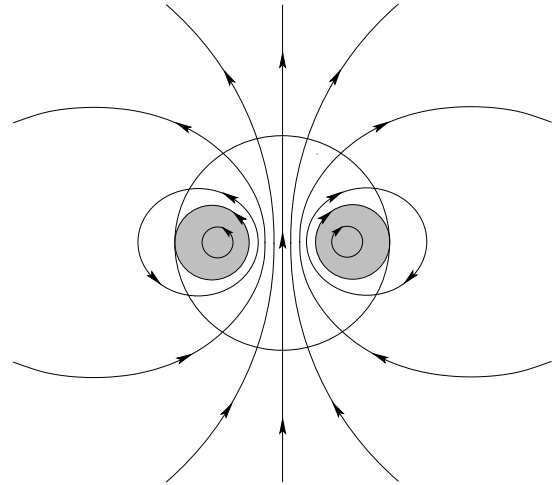}
\caption{The stable linked poloidal-toroidal torus field. Poloidal
field lines are drawn; the shaded areas represent the azimuthal field
component.}
\label{fig:stablefield}
\end{figure}

The stable field is approximately axisymmetric, and does indeed have 
both poloidal and toroidal components of comparable strength. 
The form of the resulting field is always similar: a torus-shaped configuration
embedded inside the star, with an approximately dipolar field connected to it
outside the star. Most of the field lines, projected onto a meridional plane, are 
closed inside the star, but with some extending outside it (see Fig. \ref{fig:stablefield}).

{\mk To the extent that the atmospheric field is close to a potential field, it does not support twists or internal torques.} For an axisymmetric configuration, this means that the azimuthal 
field component vanishes everywhere along a field line that extends outside the 
star. The azimuthal field component is thus confined to a torus defined by the 
lines that are closed inside the star. The field lines extending outside the star
define an approximate dipole, possibly somewhat offset from the centre. This
is in fact just what is observed in the magnetic A stars. This apparent dipole is 
not representative of the configuration as a whole: in the interior, it looks
very different. This solves one of the theoretical puzzles of the fields observed in 
A stars: they would be violently unstable if the nearly dipolar fields seen at the 
surface if were representative also of the interior.

The {\it strength} of the field found in the simulations depends on initial conditions and,
as should be expected, the orientation of the final configuration is random (at least in the 
absence of rotation).

The conclusion from these results is that an arbitrary unstable initial field does not
in general decay completely, but gets stuck in a stable equilibrium at some amplitude. 
That this should be so becomes understandable  (Moffat, 1990) in terms of the 
approximate conservation of {\it magnetic helicity}. If  $\bf A$ is a vector 
potential of the field $\bf B$, the quantity
\begin{equation}
H=\int {\bf A\cdot B} {\rm d}V,
\end{equation}
where the integral is taken over the volume inside a magnetic surface, is called the 
magnetic helicity (\cite{Woltjer:1958}). [Issues relating to the gauge used for $\bf A$ are 
involved, but of no consequence in the present context.] The helicity is conserved
under arbitrary displacements, as long as magnetic diffusion and reconnection may
be neglected. 

If an initial field configuration has a finite helicity, it can not decay further
than the lowest energy state with the same helicity (again as long as reconnection
may be ignored). In practice, reconnection cannot be ignored since diffusion will
be important on small length scales, and rapid reconnection can happen through 
regions with dynamically-generated small length scales. Nevertheless, it is found that 
in practice helicity conservation is often a good approximation on a large scale, even 
when reconnection takes place on smaller scales inside the configuration. 

{\mk A second cause for lack of conservation of helicity is the tenuous atmosphere of the 
star, which does not support significant amounts of twist in the field configuration. 
Magnetic helicity can therefore `escape' through the surface.  An example where
such a process can be observed is the field configuration in the solar corona. Most of the 
time there is little evidence of twisted or non-force-free fields, in spite of the continuous 
rearrangement of field lines at the surface caused by the magnetic activity cycle. Loss of
helicity through the surface is observed in the form of dynamic events such as 
prominence eruptions, or in a quiescent form in} large scale structures like polar crown
prominences (Low 2001, and references therein). 

{\mk We find that this process may also be important for the relaxation to the stable torus 
configuration in the present calculation and those reported in Paper I. In calculations 
where loss of magnetic field through the surface is prevented by perfectly conducting 
boundary conditions, the final field configurations were found to be noticeably different.}

{\mk In spite of these sources of helicity loss, it appears that the tendency towards 
conservation is still strong enough for a stable equilibrium to develop, at least
in many cases. In this interpretation, the orientation and strength of the final configuration 
of an unstable magnetic field in a fluid star is thus closely connected with} the helicity that 
happens to be present in the initial configuration.



The stable field produced in this way then evolves on a longer time-scale, under the 
influence of diffusion, as a sequence of quasi-static, stable equilibria. In the simulations, 
this is observed as a slow outward expansion of the configuration. In the process,
the field lines that are closed inside the star loose their azimuthal field component
as soon as they cross the surface of the star. As the azimuthal field decreases in
this way, and the poloidal component starts dominating, the configuration eventually 
becomes dynamically unstable and decays to a low value
(see paper I for further discussion).

\begin{figure*}
\includegraphics[width=0.5\hsize,angle=0]{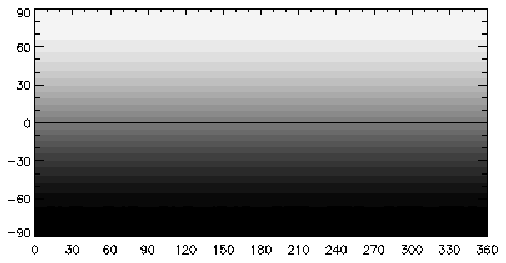}
\includegraphics[width=0.5\hsize,angle=0]{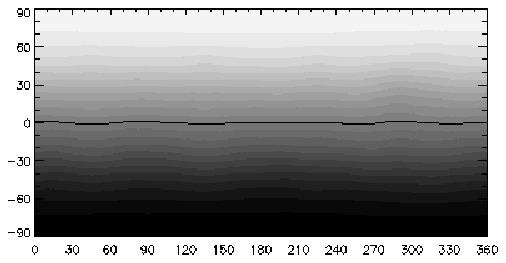}
\includegraphics[width=0.5\hsize,angle=0]{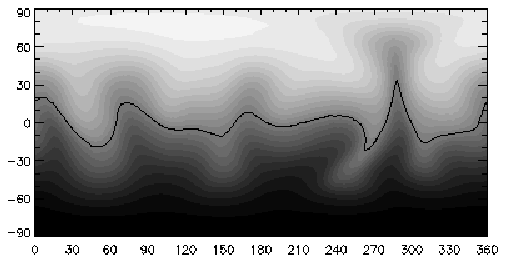}
\includegraphics[width=0.5\hsize,angle=0]{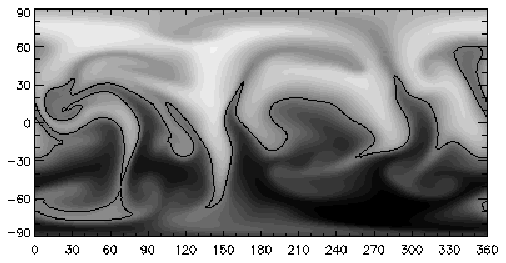}
\caption{Evolution of the magnetic field at the stellar surface. The
initial state is a uniform magnetic field in the interior and a potential field in 
the atmosphere (cf. Fig. \ref{fig:initfield}). The frames are taken at times: 
$t=0$, $4.4$, $5.0$, $5.7$ (top-left, top-right, bottom-left, bottom-right), the time unit used being the Alfv\'en
crossing time.
The radial component  $B_r$ is represented by light (positive) and dark 
(negative) shading. Also plotted, in black, is the line where $B_r=0$.}
\label{fig:ums-evol-idl}
\end{figure*}

\begin{figure*}
\includegraphics[width=0.5\hsize,angle=0]{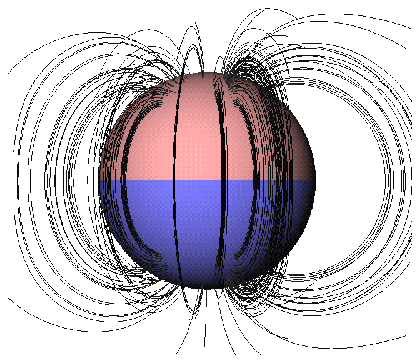}
\includegraphics[width=0.5\hsize,angle=0]{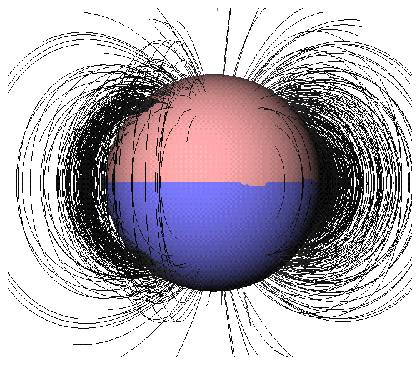}
\includegraphics[width=0.5\hsize,angle=0]{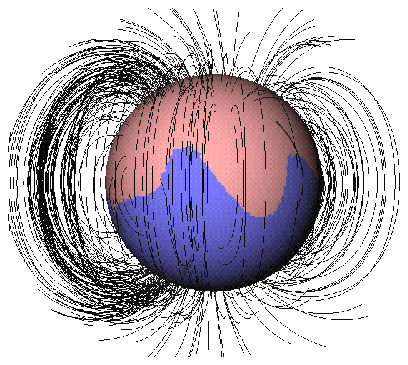}
\includegraphics[width=0.5\hsize,angle=0]{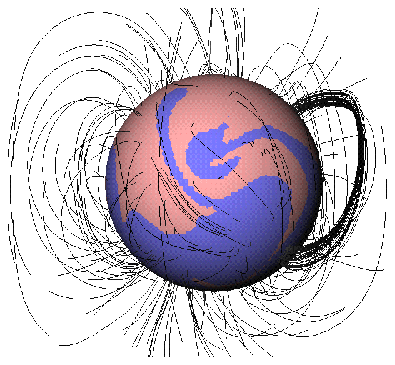}
\caption{{\mk Field lines in the stellar atmosphere. Shading on the stellar surface shows the sign of $B_r$ (light positive, dark negative). The four panels correspond to the} same times as those in Fig. \ref{fig:ums-evol-idl}.}
\label{fig:ums-evol-expb}
\end{figure*}

\section{Instability of a dipolar field}
\label{sec:instdip}

As described in Sect. \ref{sec:unstable}, a dipolar field in a ball of
conducting fluid is expected to be dynamically unstable, 
but the argument given does not say anything about how the magnetic field will evolve
in the non-linear regime. 
We use a 3D numerical simulation to see how the field evolves
under this instability, and what it evolves into.
The code used is that of Nordlund (see \cite{NorandGal:1995} for a
detailed description). The setup employed for the 
present problem is described in detail in
paper I.
In short, we model a self-gravitating ball of
perfectly conducting (up to numerical diffusion) plasma.
{\mk The initial model has the density and temperature distribution of
a polytrope of  index $n=3$. The equation of state is an ideal gas with
constant ratio of specific heats $\gamma=5/3$. The model is thus strongly
stably stratified. This model is embedded in a low-density atmosphere 
with an assigned poor conductivity, such that the field in the atmosphere
stays close to a current free configuration}, which adjusts quickly to the changing 
magnetic field distribution at the surface of the star.
We start with a field of uniform strength and direction within the 
star, and a potential field in the atmosphere.
This field is illustrated in Fig. \ref{fig:initfield}.
A spatial resolution of $144$ cubed is used here.

In principle, we could have modelled a neutron star more accurately by
using a more realistic equation of state. However, this would have
involved a fair amount of work and it was felt that the result {\mk 
 would not be affected significantly by using the equation of state of an ideal 
gas since it leads to nearly the same constraints on fluid motions.}

Since the evolution of the field advances on an Alfv\'{e}n timescale,
the evolution slows down as the field decays. 
This is impractical for numerical reasons, since the minimal time step is governed 
by the sound speed, which stays the same. The entire evolution of an unstable
magnetic configuration, however, takes place on an Alfv\'en time scale. Field
configurations which are the same except for an amplitude factor $\alpha$
will therefore evolve in the same way, with only the speed of evolution differing 
by a factor $\alpha$. Except at very high field strength, when the Alfv\'en 
speed becomes comparable to the sound speed, and on small length 
scales, where diffusion becomes important, the amplitude of the field
thus comes in only through the time-scale on which the field evolves.
We make use of this by rescaling the field strength periodically by
an overall factor, so as to keep the total field energy approximately constant.
By keeping track of the rescaling factors applied, the correct field strengths
and time axis can be reconstructed. 

Since, for numerical reasons, the magnetic energy density in the
calculations is only a factor $30$ or so smaller than the thermal
energy density, and diffusion does have an effect on small scales, some 
error is introduced by this rescaling process. Test results given in 
paper I
show that the process is accurate enough
for the present purpose, in which the final state reached by the
field configuration is of more interest than the exact dynamical
evolution to this state.

\subsection{Result}
\label{sec:fr-result}

As expected, the field is unstable and decays over a timescale of a
few Alfv\'{e}n crossing times. One of the best ways to 
visualise the results
is to look at the radial component of the field on the surface of the
star. This is plotted in Fig. \ref{fig:ums-evol-idl}, projected in a
simple manner onto two dimensions by mapping longitude to the $x$
coordinate and latitude to the $y$ coordinate. {\mk  Field lines
of the configuration are shown in} Fig. \ref{fig:ums-evol-expb}.

The initial field was of such a strength that the Alfv\'{e}n crossing
time (given by $\sqrt{4\pi\rho}R_\ast/B$, where $\rho$ and $R_\ast$
are the density and radius of the star) is roughly the same as the
temporal separation of the last three frames in Figs.
\ref{fig:ums-evol-idl} and \ref{fig:ums-evol-expb}. This confirms that the field
decays on an Alfv\'{e}n time-scale.

{\mk The decay of the field continues for the entire duration of the calculation, without any sign of levelling off.}
This can be understood in terms of the magnetic helicity introduced in Sect.
\ref{sec:evoltorus}. The initial state used in this example has zero
magnetic helicity, hence helicity conservation does not impose a lower 
limit on the final-state field. In principle, it is of course possible that
stable field configurations exist with zero total helicity. An example
could consist of two concentric tori with fields twisted in opposite directions.
Apparently, such configurations are not easily reached in the present 
case.

{\mk Another caveat concerns the degree to which helicity is actually
conserved. Since the atmosphere does not support twists in the field,
helicity can be lost through the surface of the star. Even if there is no
net helicity to being with, random fluctuations in twist propagating
through the surface might leave the interior with a finite amount of
helicity as a statistical accident of the complicated nonlinear
evolution of the field. If such a process actually takes place, it has not 
been detectable at the sensitivity of the present calculations.}

\section{Evolution of a stable twisted torus field in a neutron star
with a solid crust}
\label{sec:evoltorusns}

In earlier study of stellar magnetic fields
{\mk (paper I, see also \cite{BraandSpr:2004})}
we
found stable magnetic fields that can exist in the interior of a
star. This field is a nearly axisymmetric, twisted torus shape. We found
that such a field in an A star will slowly diffuse outwards until
becoming unstable at some point. It is not known what kind of field
may exist in a neutron star after it is formed, {\mk but  it  takes only a 
few  Alfv\'en crossing times to form the torus configurations found in
paper I. During the first minutes after the formation of the neutron star 
in a core collapse the star is still hot and completely fluid. A
magnetic field in it would relax in much the same way as in the A-star 
simulations reported in Paper I. For internal field  strengths above 
$10^{12}\,$G,  corresponding to surface dipole fields above about 
$10^{11}\,$G, the Alfv\'en crossing time would be short enough for the 
evolution to reach the stable torus equilibrium before the crust forms.} After the solid crust has formed, the
magnetic field in the fluid interior will continue to evolve under the
influence of diffusion, and stress will build up in the crust owing to
the Lorentz forces, possibly leading eventually to cracks, rather like
the processes in the Earth's crust. This buildup of stress can be 
modelled with our numerical simulations.

The mechanical properties of the crust are not well understood and no
attempt is made here to model its behaviour. We simply look at the
Lorentz forces acting on it before it cracks, not at the cracking
process itself.

The setup of the MHD code is described in Sect. \ref{sec:instdip}
and in detail in
paper I.
For this calculation we
use the twisted torus as the initial field, precisely that field which
is produced by the evolution with the same code of a random initial
field. 
Since we are only interested in what happens
inside the star and the crust, and not in what happens in the
atmosphere, we model a smaller volume than was modelled in our
previous study -- the computational box is a cube of side $2.2R_\ast$, as
opposed to the $4.5R_\ast$ used in the previous study and in the runs
described in Sect. \ref{sec:fr-result}.

To the code which we have used previously we just need to add the
solid crust. We model this as a zone between two radii in which the
velocity field is held at zero (between $r=0.86R_\ast$ and $r=1.07R_\ast$ where
$R_\ast$ is the radius of the star) and in which the magnetic field is
held constant.

\subsection{Visualising and understanding the results}

\begin{figure*}
\includegraphics[width=1.0\hsize,angle=0]{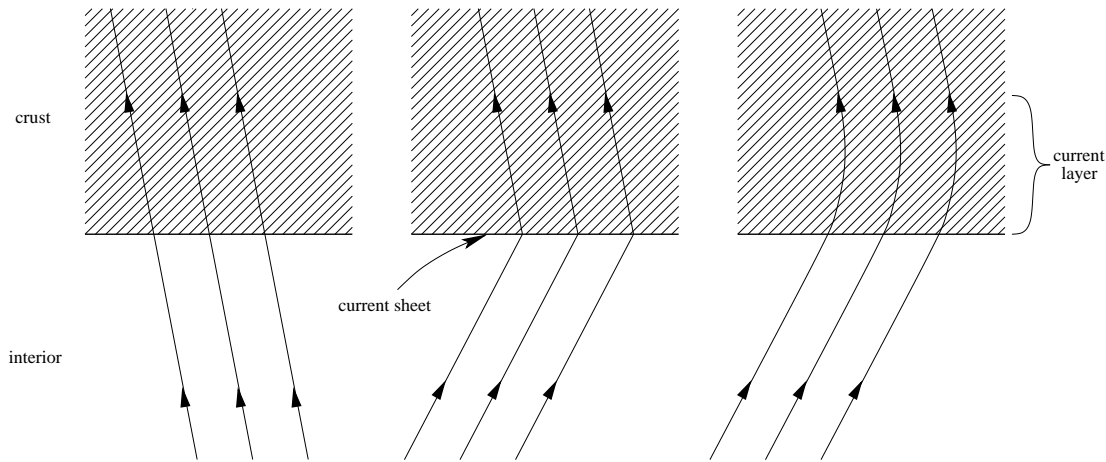}
\caption{Diagram showing the movement of field lines. On the left, the
field lines at the moment that the crust freezes. In the middle, after
the field in the interior of the star has changed, the perfectly
conducting crust carries a surface current where the Lorentz forces
act. On the right, the crust is not perfectly conducting, having
magnetic diffusivity $\eta$ and contains, after time $t$,
a thin conducting layer (of thickness of order $\sqrt{t\eta}$).}
\label{fig:layers}
\end{figure*}

We are wanting to look at the stresses that build up in the solid
crust, as these will eventually cause the crust to crack. In a real
star with an infinitely conducting crust and a diffusive fluid
interior, we would expect to see a discontinuity in the field at the
lower boundary of the crust, and a current sheet. In a more realistic
crust of high but finite conductivity, the field lines would bend
within a thin layer at the bottom of the crust, without any
discontinuity in the magnetic field. The Lorentz force created by the
change in the field will act on the crust in this thin layer. This is
illustrated in Fig. \ref{fig:layers}.

At the upper boundary we would not expect
to see any significant discontinuity since the field outside is
essentially a potential field and will consequently remain constant
after the crust has formed.

An MHD code of the type used here cannot model surface currents, yet
we are trying to model a crust of infinite conductivity, in which the
magnetic field is held constant. In order to look at the stress acting
on the crust, therefore, we need to integrate the Lorentz force over a layer of at least
a few grid spacings either side of the lower boundary of the
crust. But what kind of Lorentz force are we interested in?

At the moment when the crust freezes, there will already be a non-zero
Lorentz force. If we express this force as the divergence of a scalar
plus the curl of a vector, the former will already be balanced by
pressure (buoyancy) forces. The latter will give rise to fluid
currents which will then have the effect of reducing the driving force
itself -- we can therefore expect this component of the force to be
small at the moment when the crust freezes. To look at the stresses on
the crust, therefore, we need to subtract from the Lorentz force that
which was present at the moment of freezing.

If the field is of order $10^{15}$ gauss, a typical length scale
$\mathcal{L}$ is
$1$ km and the crust has a density of $10^{14}\,\rm{g\,cm}^{-3}$ then the
Lorentz force per unit mass will be around
$B^2/4\pi\mathcal{L}\rho\sim10^8\,\rm{cm\,s}^{-2}$. The force of
gravity per unit mass will be $GM/R^2\approx 2\times10^{14}\,\rm{cm\,
s}^{-2}$, which is balanced by pressure. The vertical component of the
Lorentz force can
therefore be balanced by the change in gravitational force brought
about by a vertical displacement of just $0.3$ cm.

We feel it is safe to assume that the crust is flexible enough to
accommodate vertical movements of this size without cracking. The
vertical component of the Lorentz force, therefore, is not of any
great interest to us. Of the remaining Lorentz force, that is, the horizontal
component
minus that at the moment of crust formation, it is only the
part which can be expressed as the curl of a vector normal to the
surface which interests us, as the other part can be supported by
pressure forces. This is the part of the Lorentz force given
by

\begin{eqnarray}
F_x & = & \frac{1}{4\pi} \frac{\partial(B_z B_x)}{\partial z}\qquad \rm{and}
\nonumber\\
F_y & = & \frac{1}{4\pi} \frac{\partial(B_z B_y)}{\partial z}.
\end{eqnarray}

This is then integrated over a zone from $r=0.79R_\ast$ to $r=0.93R_\ast$ (the
inner boundary of the crust is at $r=0.86R_\ast$), and can be plotted as
arrows on a 2-D projection of the star. The projection used is that
where longitude forms the x-axis and latitude forms the y-axis; to do
this an axis is of course required -- we use the dipole axis of the
radial component $B_r$ of the field on the surface.

It is also useful to calculate an average Lorentz stress over the
whole of the crust, to see how it changes with time. This is done by
simply taking a root-mean-square of the component described above.

\subsection{Result}

\begin{figure}
\includegraphics[width=1.0\hsize,angle=0]{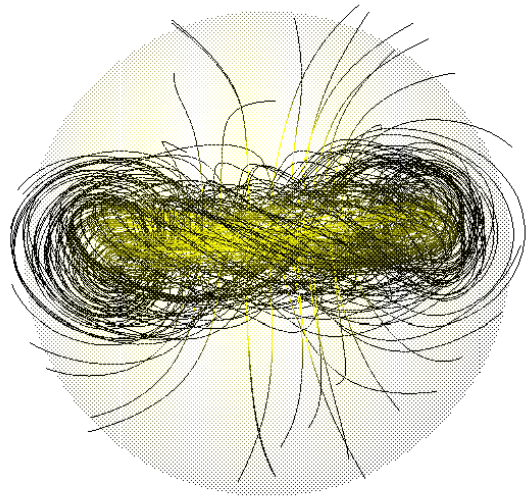}
\caption{{\mk Field lines at the beginning of the
run, with the core of the torus configuration shown as a shaded surface.} A transparent sphere representing the surface of the star is also plotted.
\label{fig:initfield-side}}
\end{figure}

\begin{figure}
\includegraphics[width=1.0\hsize,angle=0]{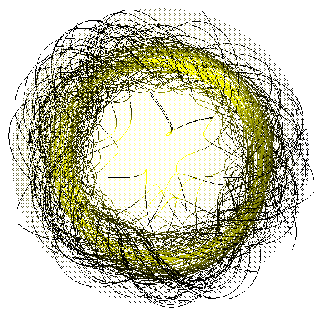}
\caption{As Fig. \ref{fig:initfield-side}, but viewed from above.}
\label{fig:initfield-top}
\end{figure}

The evolution of the field inside a star with a solid crust was
modelled at a resolution of $96^3$, producing a crust $9$ grid
spacings thick. {\mk The code runs for about 20 Alfv\'{e}n
crossing times, corresponding to about 1/4 of a diffusion time
$R^2/\eta$, where $R$ is the radius of the star. Since the
characteristic length scale of the torus configuration is about
3 times smaller (cf. Figs \ref{fig:initfield-side} and \ref{fig:initfield-top}),
this corresponds to about 2 times the characteristic diffusion
time of the torus configuration itself. After this time, the code
crashes when the stresses at the surface become too large. 
This limitation} does not matter to us, since we are only interested in
looking at the buildup of stresses and not in the cracking process
itself.

\begin{figure}
\includegraphics[width=1.0\hsize,angle=0]{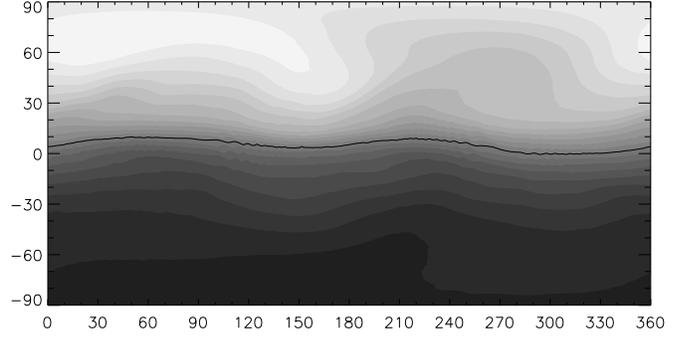}
\caption{{\mk Initial condition for the magnetar field evolution calculation,
showing the radial component $B_r$ of the magnetic field on the surface of 
the star, in longitude (horizontal) and latitude. Light (dark) shows
positive (negative) $B_r$. The black line shows  the magnetic equator 
$B_r=0$.}}
\label{fig:br}
\end{figure}

\begin{figure}
\includegraphics[width=1.0\hsize,angle=0]{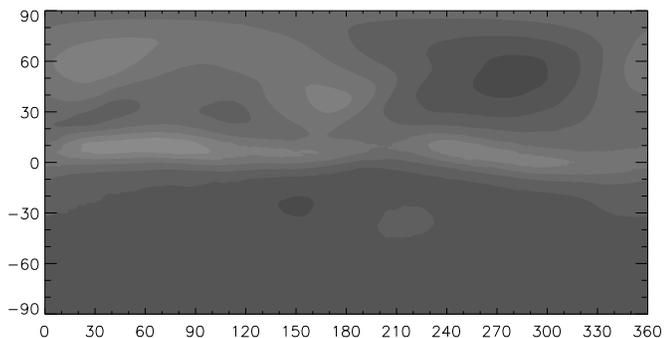}
\caption{{\mk As Fig.\ \ref{fig:br}, showing the modulus of the magnetic field 
$|\mathbf{B}|$ on the surface of the star. Light (dark) represent regions 
of strong (weak) field}.}
\label{fig:bmod}
\end{figure}

The initial conditions for the run are taken from the output from an
earlier run in which an initial random field configuration evolved
into a twisted torus 
(see paper I).
The radial component
of the field on the surface is shown in Fig. \ref{fig:br}. It can be
seen that the field is chiefly dipole in nature. Fig. \ref{fig:bmod}
shows the modulus of the field, which is strongest just above the
magnetic equator -- the dipole field is slightly offset from the
centre of the star.

{\mk Field line tracings are shown in} Figs.~
\ref{fig:initfield-side} and \ref{fig:initfield-top}. The twisted
torus shape is clearly visible.

\begin{figure}
\includegraphics[width=1.0\hsize,angle=0]{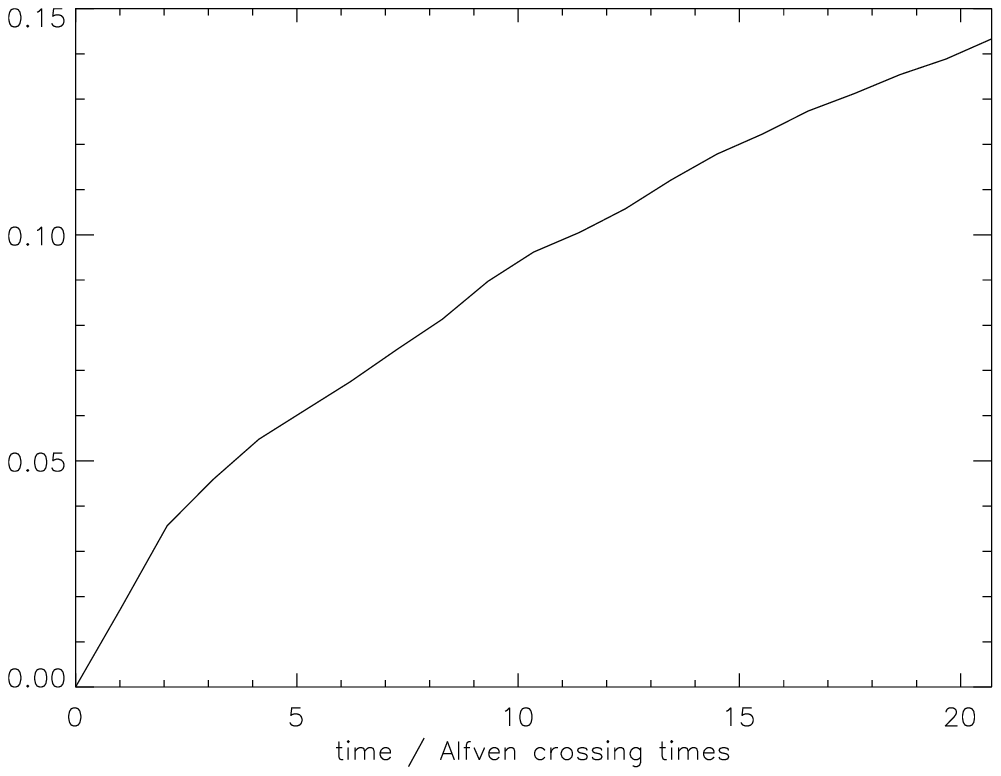}
\caption{Root-mean-square stress as a function of
time. The unit used is the mean magnetic pressure $B^2/8\pi$ in the
stellar interior at $t=0$.}
\label{fig:meanstress}
\end{figure}

Fig.~\ref{fig:meanstress} {\mk shows the} r.m.s. stress per unit area
as a function of time.

\begin{figure*}
\includegraphics[width=0.5\hsize,angle=0]{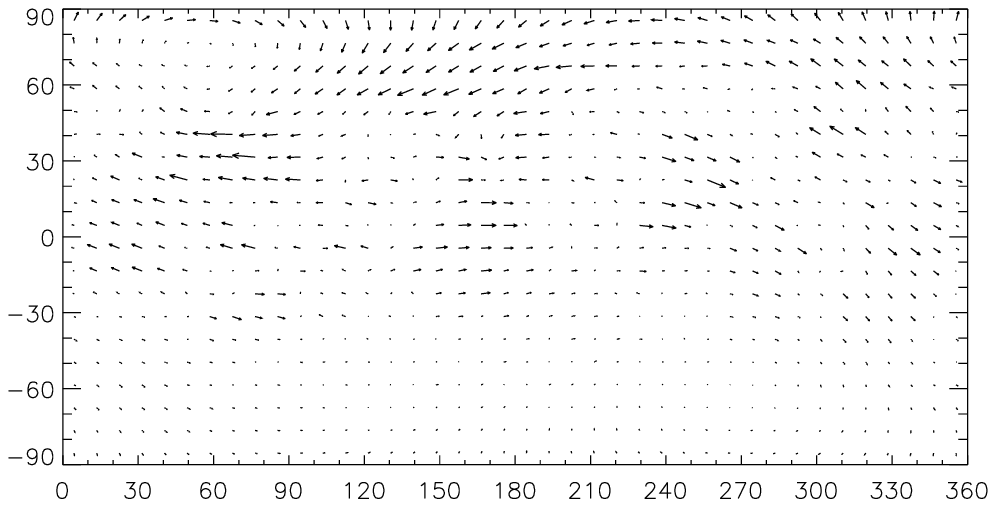}
\includegraphics[width=0.5\hsize,angle=0]{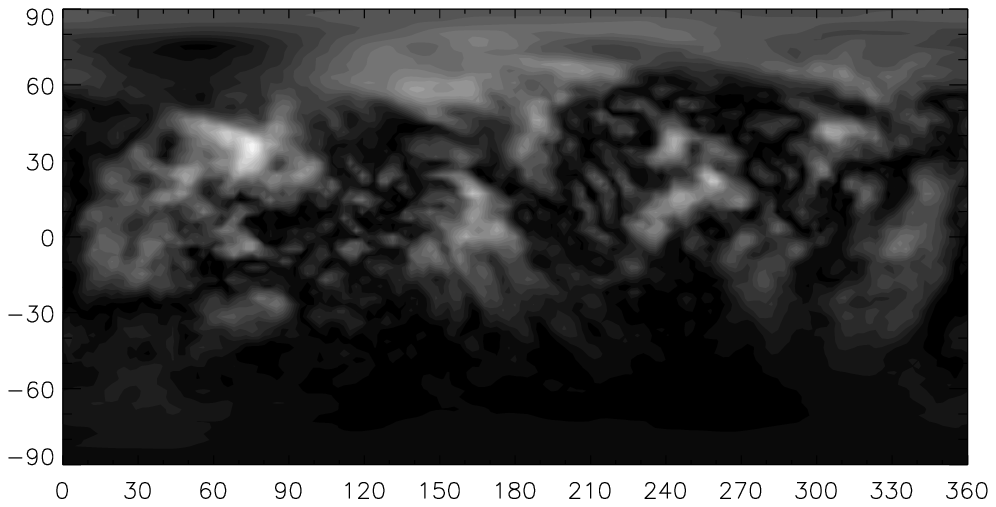}
\includegraphics[width=0.5\hsize,angle=0]{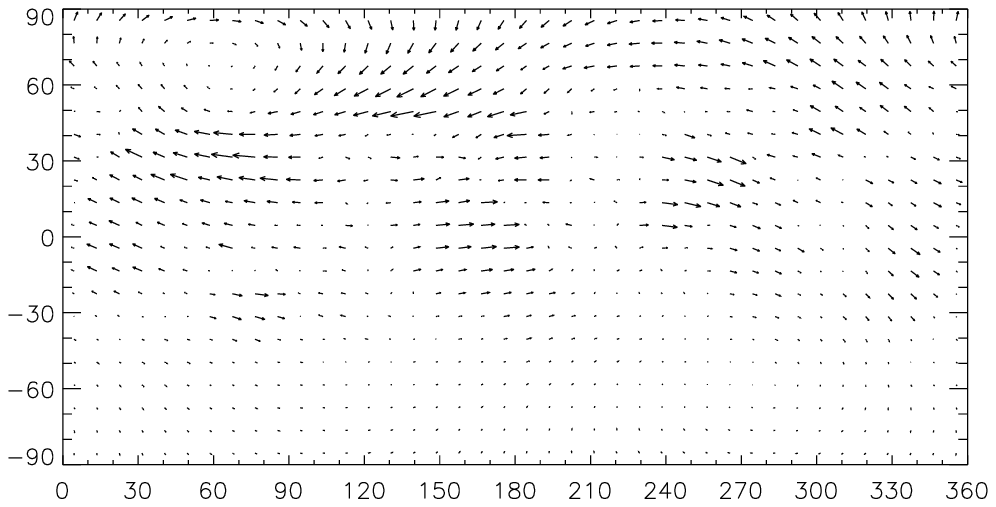}
\includegraphics[width=0.5\hsize,angle=0]{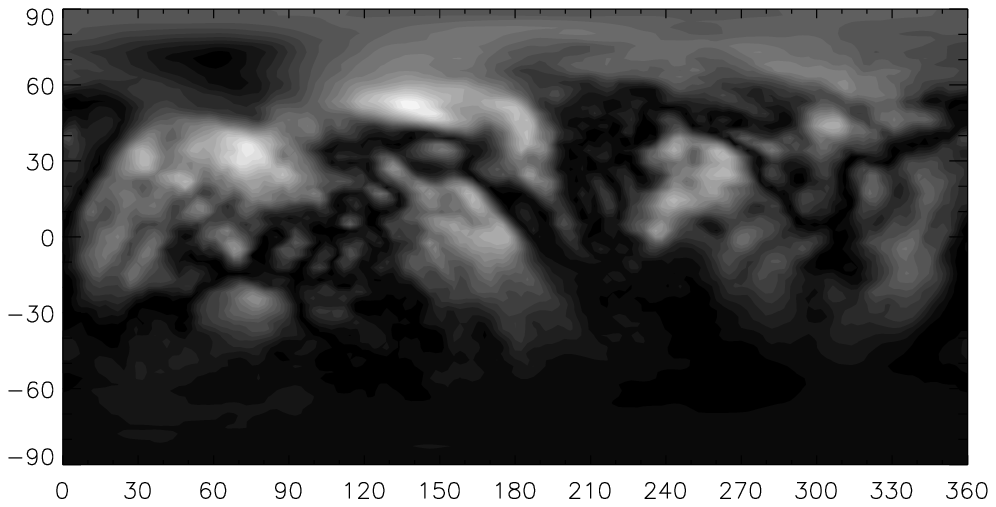}
\includegraphics[width=0.5\hsize,angle=0]{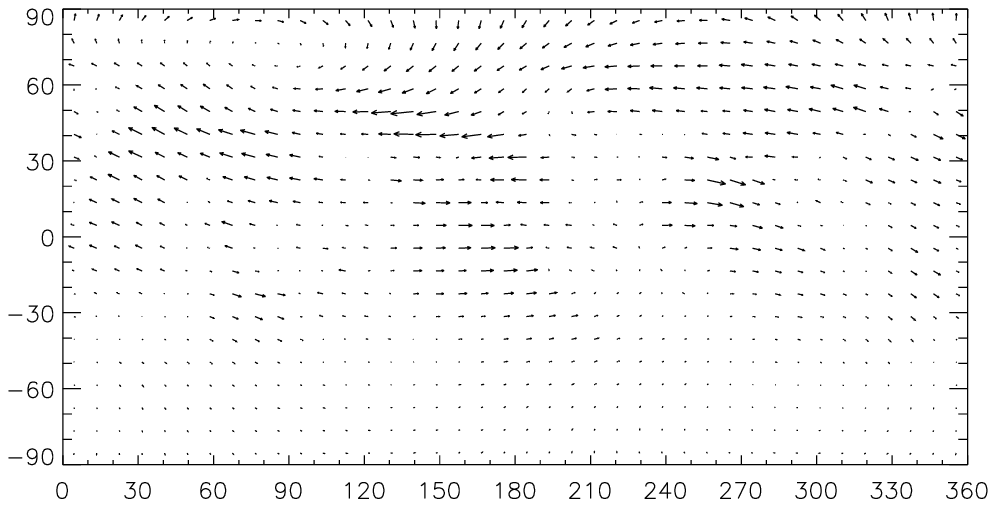}
\includegraphics[width=0.5\hsize,angle=0]{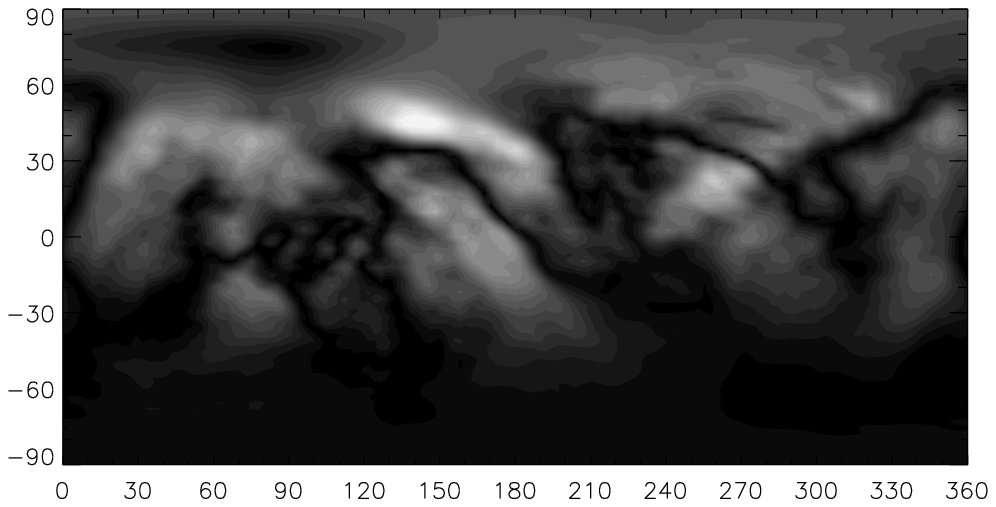}
\includegraphics[width=0.5\hsize,angle=0]{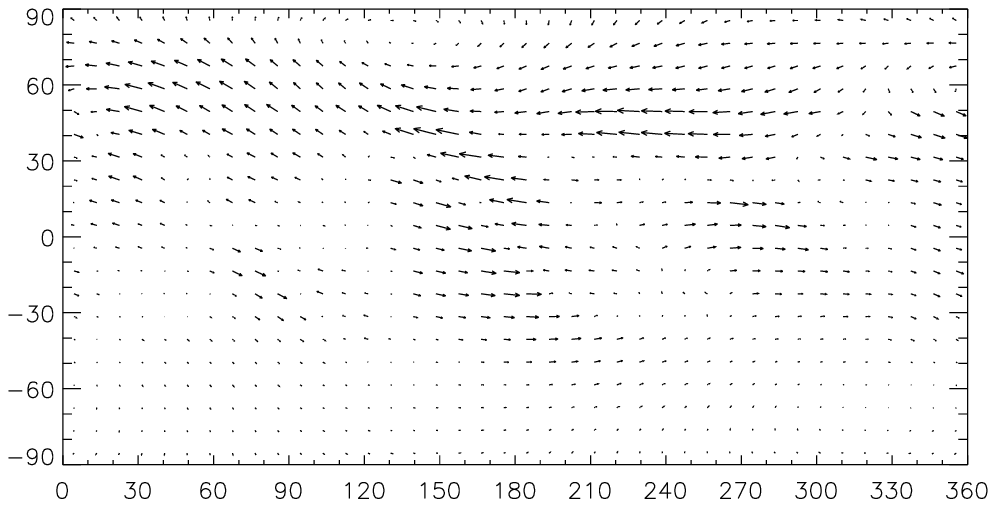}
\includegraphics[width=0.5\hsize,angle=0]{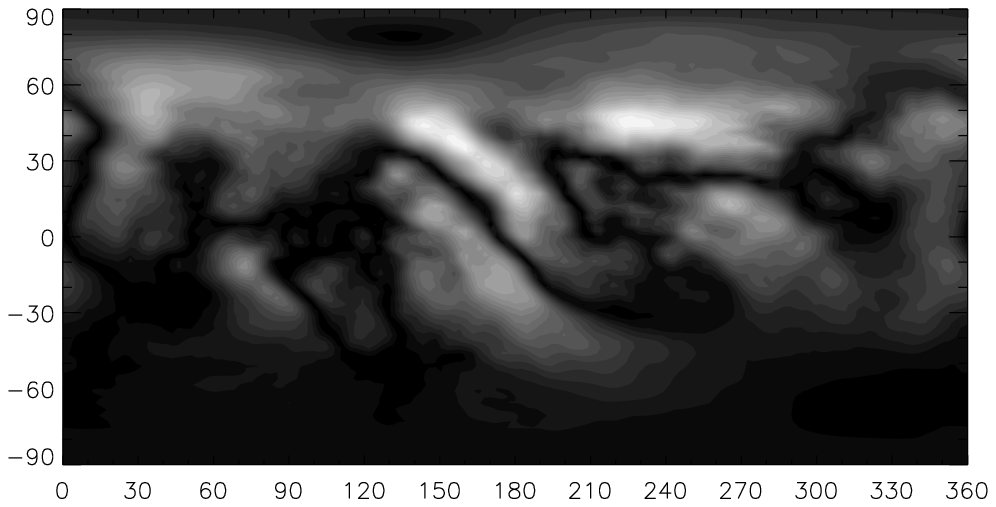}
\caption{Maps of stress {\mk at the outer surface}, at four roughly equal
temporal intervals: $t=5.3, 10.4, 15.6, 20.7$ Alfv\'en crossing-times respectively from top
to bottom. On the left hand side, {\mk arrows show the direction and strength of the Lorentz stress. The right hand side shows the magnitude of the Lorentz stress. (Arrows and grey scales are scaled to maximum over the surface in each panel.)}}
\label{fig:mapofstress}
\end{figure*}

Fig. \ref{fig:mapofstress} {\mk shows the horizontal components of the stress at the surface} (as
described in the previous section) at four times after the {\mk assumed formation} of
the crust at $t=0$, at roughly equal intervals.
It can be seen that there are lines separating areas with their
Lorentz stress in opposite directions (these show up as black lines
between the bright areas on
the right-hand-side of Fig. \ref{fig:mapofstress}; they are mainly
present in the equatorial zone). It is presumably along these
lines that the fault lines will appear. There are also patterns where
the Lorentz force is exerting a torque on a part of the the crust;
this is visible mainly in the polar regions. {\mk Smaller length scales were
still present in the initial configuration at a low amplitude; their more rapid 
diffusion causes the smaller scales seen in the stress pattern in Fig.  
\ref{fig:mapofstress}, as well as in the somewhat faster initial development 
of  the stress in Fig. \ref{fig:meanstress}. }

\section{Application: scenario for magnetar evolution}
\label{sec:evolmag}

The results of this study lead logically to a model for magnetar
evolution. First, we require the crust to have a much higher
electrical conductivity than the interior, so the field is `frozen'
into the crust while the field continues to evolve in the fluid interior. The magnetic
diffusivity $\eta$ of the interior has to be such that the timescale of the
field's evolution, $\mathcal{L}^2/\eta$ (where $\mathcal{L}$ is a
typical length scale), is of the order of $10^4$ years, the observed
lifetime of these objects. The second requirement of the model is that
the star contains a stable field at the moment when the crust starts
to form, of the order of $10^{15}$ gauss.

The hypothesis then is that this stable field in the fluid interior
evolves under the influence of diffusion, while the field in the crust
is held relatively constant by higher conductivity. This causes stress
to build up in the lower part of the crust, which at some point
becomes strong enough to cause fractures to appear, and energy is
released by two mechanisms. In the movement of the crust, magnetic
energy is converted to kinetic energy and then to heat. In the
atmosphere, the movement of different parts of the crust causes the
field lines to become twisted, giving rise to large currents in the
tenuous medium. This results in Ohmic heating and eventual relaxation
back to a potential field.

\section{Summary}
\label{sec:summary}

We have studied the release of magnetic energy in
a strongly magnetised neutron star, with a view to finding an
explanation for the high luminosity of the soft gamma repeaters and
anomalous X-ray pulsars. These objects are observed to release energy
over a timescale of around $10^4$ years, much greater than the
Alfv\'en timescale (the time taken for an Alfv\'en wave to travel
across the star, $0.1$ s) on which an unstable field
evolves. This leads us to the conclusion that the magnetic field must
go through a sequence of quasi-static, stable equilibria,
possibly occasionally punctuated by instability. A central question is
thus what this stable magnetic equilibrium looks like and how it 
can form.

By following the decay of arbitrary initial field configurations with 3D 
magnetohydrodynamic simulations we have shown in
Paper I
that a stable configuration exists in the form
of a torus of linked poloidal-toroidal field, {\mk confirming educated guesses existing in the  
literature (e.g. Prendergast 1956). We found there that in} all cases, the field
either reaches this torus field, or it decays to nothing; no other end-states
were found. Whether the field evolves into the stable torus shape was 
found to depend on how concentrated the field is in the centre of the
star. A field concentrated in the stellar core can reach the torus
configuration, whilst a field whose energy distribution extends to
greater radii and into the atmosphere usually decays to nothing.

This result can be understood in terms of helicity conservation. In the 
absence of reconnection, helicity is a conserved quantity, so that an 
arbitrary field will evolve to the lowest energy state with the same 
helicity. This appears to be the linked poloidal-toroidal torus. If the 
initial field is buried less deeply in the star and has too strong a contact 
with the atmosphere, an environment where reconnection cannot be 
ignored, helicity conservation is no longer valid and both the helicity and
field energy can decay to zero.

A configuration considered before in connection with the stability of
magnetic fields in neutron stars is {\mk a uniform field inside matching to a 
pure dipole outside the star. The instability argument of Flowers and 
Ruderman (1977) shows that the energy of this configuration is reduced by
rotating one half of the star by 180$^\circ$ with respect to the other
along a magnetic surface. Our numerical simulation of this configuration
confirms its instability, but instead of a large scale rotation, the instability
is found to take place at higher azimuthal wavenumbers, with displacements
reminiscent of other interchange instabilities such as Rayleigh-Taylor.}

The Flowers-Ruderman field has zero helicity, hence, by a symmetry
argument we {\mk may} expect that it will not develop the stable tori
found in Paper I.
A torus must have an axis, which can only be parallel to the original
axis; the toroidal component must then go around this axis in one
direction or the other, but there is no way to decide which direction.
However, this 
argument does not entirely rule out the existence of stable fields with 
zero helicity as end states of unstable evolution. For instance two tori 
on top of each other, with opposite `handedness'. {\mk It is also possible
that the interior could acquire some helicity by loss of helicity through 
the surface of the star in the course of the evolution of the field . We 
have seen no evidence of this occurring, however, and 
our tentative conclusion is that the decay of a Flowers-
Ruderman configuration may well be rather complete.}

We have studied the slow evolution of {\mk a stable torus} configuration 
in the diffusive interior of a neutron star, underneath a solid and
perfectly-conducting crust. As the field changes under the influence
of diffusion, stress develops in the crust. We find that this stress
is dominated by patterns that would cause strong rotational
displacements on the neutron star surface when released, consistent
with the model of SGR outbursts developed by Thompson \& Duncan 
(1995, 2001). 

{\mk One may speculate what the final evolution of a magnetar field
would be as this process continues. In Paper I we found that the diffusing
torus gradually looses the azimuthal field component that provides
stability to the torus, by loss through the surface of the star. At some
point, this configuration, dominated by the poloidal component,
became unstable in the way predicted by the Flowers-Ruderman
argument, developing an $m=2$ distortion. The further evolution
of this configuration led to a final episode of rapid decay. If this
also happens in magnetars, it predicts a final episode of rapid decay,
perhaps in the form of a giant outburst, as suggested before by 
Eichler (2002).	}

\begin{acknowledgements}
{\mk We thank the referee for his suggestions and valuable criticism of 
an earlier version of the text.}
\end{acknowledgements}

\begin{appendix}

\end{appendix}


\begin{flushleft}

\end{flushleft}


\begin{thebibliography}{99}

\bibitem[Acheson 1978]{Acheson:1978}Acheson, D.J., 1978, Phil.
Trans. Roy. Soc. Lond., A289, 459.
\bibitem[Blandford, Applegate, \& Hernquist(1983)]{1983MNRAS.204.1025B} 
Blandford, R.~D., Applegate, J.~H. and Hernquist, L.\ 1983, \mnras, 204, 
1025.
\bibitem[Braithwaite 2005]{Braithwaite:2005}Braithwaite,
J., 2005, to be published.
\bibitem[Braithwaite \& Nordlund 2005]{BraandNor:2005}Braithwaite,
J. and Nordlund, \AA, 2005, to be published. (Paper I).
\bibitem[Braithwaite \& Spruit 2004]{BraandSpr:2004}Braithwaite,
J. and Spruit, H.C., 2004, Nature, 431, 891..
\bibitem[Cline 1982]{Cline:1982}Cline, T.L., 1982, Gamma ray
transients and related astrophysical phenomena, AIP,
eds. Lingenfelter et al., p17.
\bibitem[Duncan \& Thompson 1992]{DunandTho:1992}Duncan, R.C. and
Thompson, C., 1992, ApJ, 392, L9.
\bibitem[Eichler 2002]{Eichler:2002}Eichler, D.\  2002, MNRAS, 335, 883.
\bibitem[Flowers \& Ruderman 1977]{FloandRud:1977}Flowers, E. and
Ruderman, M.A, 1977, ApJ, 215, 302.
\bibitem[Hurley et al. 1999]{Hurleyetal:1999}Hurley, K. et al., 1999,
ApJ, 510, L107.
\bibitem[Kouveliotou et al. 1998]{Kouveliotouetal:1998}Kouveliotou,
C. et al., 1998, Nature, 393, 235.
\bibitem[Low 2001]{Low:2001}Low, B.C., 2001, JGR, 106, 25141.
\bibitem[Markey \& Tayler 1973]{MarandTay:1973}Markey, P. and Tayler,
R.J., 1973, MNRAS, 163, 77.
\bibitem[Markey \& Tayler 1974]{MarandTay:1974}Markey, P. and Tayler,
R.J., 1974, MNRAS, 168, 505.
\bibitem[Mazets et al. 1999]{Mazetsetal:1999}Mazets, E.P. et al.,
1999, Astronomy letters, 25, 635.
\bibitem[Mereghetti 2000]{Mereghetti:2000}Mereghetti, S., 2000, in The
neutron star -- black hole connection, eds. Kouveliotou, Paradijs \&
Ventura, p. 351.
\bibitem{}Moffat, K.H., 1990, private communication.
\bibitem[Nordlund \& Galsgaard 1995]{NorandGal:1995}Nordlund, \AA. and
Galsgaard, K., 1995, \\http://www.astro.ku.dk/$\sim$aake/papers/95.ps.gz
\bibitem[Palmer et al. 2005]{Palmeretal:2005}Palmer, D.~M., Barthelmy, S., Gehrels, N., et al.\  2005, Nature, 434, 1107. 
\bibitem[Pitts \& Tayler 1986]{PitandTay:1986}Pitts, E. and Tayler,
R.J., 1986, MNRAS, 216, 139.
\bibitem[Prendergast 1956]{Prendergast:1956}Prendergast, K.H., 1956,
ApJ, 123, 498.
\bibitem[Rothschild et al. 1994]{Rothetal:1994}Rothschild, R.E.,
Kulkarni, S.R. and Lingenfelter, R.E., 1994, Nature, 368, 432.
\bibitem[Shapiro \& Keukolsky 1983]{ShaandTeu:1983}Shapiro, S.L. and
Teukolsky, S.A., 1983, ``Black holes, white dwarfs and neutrons stars:
the physics of compact objects'', John Wiley \& Sons.
\bibitem[Spruit 2002]{Spruit:2002}Spruit, H. C., 2002, A\&A, 381, 923.
\bibitem[Tayler 1973]{Tayler:1973}Tayler, R.J., 1973, MNRAS, 161, 365.
\bibitem[Thompson \& Duncan 1993]{ThoandDun:1993}Thompson, C. and
Duncan, R.C., 1993, ApJ, 408, 194.
\bibitem[Thompson \& Duncan 1995]{ThoandDun:1995}Thompson, C. and
Duncan, R.C., 1995, MNRAS, 275, 255.
\bibitem[Thompson \& Duncan 1996]{ThoandDun:1996}Thompson, C. and
Duncan, R.C., 1996, ApJ, 473, 322.
\bibitem[Thompson \& Duncan 2001]{ThoandDun:2001}Thompson, C. and
Duncan, R.C., 2001, ApJ, 561, 980.
\bibitem[Woltjer 1958]{Woltjer:1958}Woltjer, L., 1958,
Proc. Nat. Acad. Sci., 44, 489.
\bibitem[Woods et al. 2000]{Woodsetal:2000}Woods, P.M. et al., 2000,
ApJ, 535, L55. 
\bibitem[Wright 1973]{Wright:1973}Wright, G.A.E., 1973, MNRAS, 162, 339.
\bibitem[Zhang \& Low 2003]{ZhaandLow:2003}Zhang, M. and Low, B.C.,
2003, ApJ, 584, 479.
Press).


\end{thebibliography}
\end{document}